\documentstyle[12pt,psfig]{article}
\begin{document}
\baselineskip=8 true pt
\vskip 0.2in
\begin{center}
{\Large {Liquid-gas phase transition in finite nuclei}} 
\vskip 0.2in
{\normalsize  J. N. De$^{1)}$, S. K. Samaddar$^{2)}$ and 
S. Shlomo$^{3)}$}
\end{center}
\vskip 0.5cm
\noindent
{\small {\it 1) Variable Energy Cyclotron Centre,
     1/AF Bidhan Nagar, Calcutta 700 064, India}}
\vskip 0.5cm
\noindent
{\small {\it 2) Saha Institute of Nuclear Physics,
           1/AF Bidhan Nagar, Calcutta 700 064,
           India}}
\vskip 0.5cm
\noindent
{\small {\it 3) Cyclotron Institute, Texas A\&M University,
College Station, TX 77843-3366, USA}}
\parindent=20pt
\vskip 0.4in
\noindent
\begin{abstract}
In a finite temperature Thomas-Fermi framework, we calculate density  
distributions of hot nuclei enclosed in a freeze-out volume of few
times the normal nuclear volume and then construct the caloric curve,
with and without inclusion of radial collective flow. In both cases,
the calculated specific heats $C_v$ show a peaked structure signalling
a liquid-gas phase transition. Without flow, the caloric curve indicates
a continuous phase transition whereas with inclusion of flow, the
transition is very sharp. In the latter case, the nucleus undergoes a
shape change to a bubble from a diffuse sphere at the transition
temperature.  
\end{abstract}
\medskip

1. INTRODUCTION
\vskip 0.5cm
The equation of state (EOS) of nuclear matter 
constructed with realistic effective interactions \cite{BSSD}
shows that 
the nuclear liquid coexists in phase-equilibrium with the      
vapour surrounding it below a critical temperature $\approx 15-20$ MeV.
Searching for signals for liquid-gas phase transition in nuclear
matter or in finite nuclei has become one of the fundamental issues in
medium and higher energy nuclear collisions in recent times after the
initial experimental observations of a power law behaviour in the mass or
charge distribution of nuclear fragments in proton and heavy ion induced
reactions which is usually taken to be indicative of a critical behaviour
\cite{FIS}. On the other hand, different models of liquid-gas phase
equilibrium in nuclei \cite{BSSD,LB} show that there is a limiting temperature
beyond which the nucleus is unstable and the nuclear liquid and the
surrounding gas can not coexist in thermodynamic equilibrium. The limitation
in temperature also seems to be an experimentally supported observation
\cite{NAT}. A possible connection between the phase transition temperature
and the limiting temperature is therefore called for. Normally phase
transitions are signalled by peaks in the specific heat $C_v$ at constant
volume as temperature increases. The microcanonical algorithm of Gross
\cite{GRO} designed for multifragmentation calculations show such a
peaked structure for $C_v$, so also the calculations by the Copenhagen
school \cite{BON} in a canonical description. Recent calculations by
Dasgupta et al. \cite{DAS} in the lattice gas model for fragmentation also
show such a peak for the specific heat. These calculations however do not
allow for any detailed study of the changes in the density distribution
at criticality.
\newpage
Renewed interest in the liquid-gas phase transition has been fueled further by
the recent experimental observation \cite{POC} that in Au+Au collisions
at 600A MeV, the temperature in the caloric curve remains almost constant
in the excitation energy range of $\approx 4-10$ MeV per nucleon. This
is reminiscent of a first order phase transition and the width in
the excitation energy at constant temperature may be called the latent
heat for liquid-vapour phase transition. Another set of experimental data
with Au on C is reported recently \cite{HAU} at a little higher energy
of 1A GeV where the caloric curve looks somewhat different; here around
6-7 MeV the temperature rises slowly but steadily with excitation energy
indicative of a continuous phase transition. These two different behaviours
in the caloric curve possibly reflect subtle changes in the physical
process as the bombarding energy increases or as the colliding mass
changes. 
In this paper, we propose
an explanation for them in a finite temperature Thomas-Fermi theory 
(FTTF). Part of the work has been published recently \cite{DDSS}.
\vskip 0.5cm
2. THEORETICAL FRAMEWORK
\vskip 0.5cm
In the following, we give a brief outline for determining the density
profiles of hot nuclei in the FTTF approximation starting from an effective
nucleon-nucleon interaction.
\vskip 0.5cm
{\bf 2.1. The interaction}
\vskip 0.5cm
The energy density of the nucleus is calculated with a Seyler-Blanchard type
momentum and density dependent finite range two-body effective interaction
\cite{BSSD}. It is given by
$$
v_{{eff}}(r,p,\rho)=C_{l,u} [v_{1}(r,p)+v_{2}(r,\rho)]
\eqno (1)
$$
$$
v_{1}=-(1-p^2/b^2)f({\vec r_{1}},{\vec r_{2}}),
$$
$$ 
v_{2}=d^2[\rho_1(r_1)+\rho_2(r_2)]^n f({\vec r_1},{\vec r_2})
\eqno (2)
$$
with
$$
f({\vec r_1},{\vec r_2})=\frac{e^{- \mid {\vec r_1}-{\vec r_2} \mid } /a}
{ \mid {\vec r_1}-{\vec r_2} \mid /a}.
\eqno (3)
$$
Here $a$ is the range of the interaction, $b$ denotes its strength of
repulsion in the momentum dependence, $r=|{\vec r_1}-{\vec r_2}|$ and
$p=|{\vec p_1}-{\vec p_2}|$ are the relative distance and momenta of
the two interacting nucleons. The subscripts $l$ and $u$ in the strength
$C$ refer to like pair (n-n or p-p) or unlike pair (n-p) interaction
respectively, $d$ and $n$ are measures of the strength of the density           
dependence of the interaction and $\rho_1$ and $\rho_2$ are the densities at
the sites of the two nucleons. The values of the potential parameters
are given in \cite{DE}.
\medskip

{\bf 2.2. Self-consistent density profile and the caloric curve}
\vskip 0.5cm
The interaction energy density for a finite nucleus with the interaction
chosen is given by
$$
\varepsilon_{I}(r)=\frac {2}{h^3}\sum_{\tau}[\frac{1}{h^3}
\int \{ v_1(|{\vec r-\vec r^{\prime}}|,|{\vec p-\vec p^{\prime}}|)+
v_2(|{\vec r-\vec r^{\prime}}|,\rho)\}\,\times 
$$
$$
\{C_ln_\tau({\vec r^{\prime}},
{\vec p^{\prime}}) + C_{u}n_{-\tau}({\vec r^{\prime}},{\vec p^{\prime}})\}
n_{\tau}({\vec r},{\vec p})]\,d{\vec r^{\prime}}\,d{\vec p}\,               
d{\vec p^{\prime}},
\eqno (4)
$$
and the kinetic energy density is given by
$$
{\cal K}(r)=\frac{2}{h^3}\sum_{\tau}\int\frac{p^2}{2m_\tau}n_{\tau}(
{\vec r},{\vec p})d{\vec p}.
\eqno (5)
$$
Here $\tau$ is the isospin index, $n_{\tau}({\vec r},{\vec p})$ the occupation           
probability and $m_{\tau}$ the nucleon mass. The coulomb interaction energy
density is taken to be the sum of the direct and exchange term. 

The occupation probability is obtained by minimising the thermodynamic
potential
$$
G=E-TS-\sum_{\tau}\mu_{\tau}A_{\tau},
\eqno (6)
$$
where E and S are total energy and entropy of the system at temperature T
and $A_{\tau}$ the number of neutrons or protons, $\mu_{\tau}$ being their
chemical potential. In the local density approximation, the occupation
probability can be written as
$$
n_{\tau}({\vec r},{\vec p})=[1+exp\{(\frac{p^2}{2m_{\tau}^{\star}(r)} +
V_{\tau}^{0}(r) + V_{\tau}^{2}(r) + \delta_{\tau , p}V_c(r)-\mu _{\tau})
/T\}]^{-1},
\eqno (7)
$$
where $m_{\tau}^{\star}(r)$ is the effective mass dependent on the
momentum dependent part of the single particle potential $p^2V_{\tau}^1(r)$
and $V_{\tau}^0,V_{\tau}^1,V_{\tau}^2,V_c$ etc are potential terms
expressions of which in terms of density are given in Ref.\cite{DE}.
The density is the momentum integral of the occupancy.
The total energy density is then written as
$$
\varepsilon(r)= \sum_{\tau} \rho_{\tau}(r)[T J_{3/2}(\eta_{\tau}(r))/
J_{1/2}(\eta_{\tau}(r))(1-m_{\tau}^{\star}(r)V_{\tau}^1(r)) +
\frac{1}{2}(V_{\tau}^0 +\delta_{\tau p}V_c(r))] ,
\eqno (8)
$$
where $J_K$ are the Fermi integrals and the fugacity $\eta_{\tau}(r)$ is
defined as 
$$
\eta_{\tau}(r)= [\mu_{\tau}- V_{\tau}^0(r) - V_{\tau}^2(r) -
\delta_{\tau p}V_c(r)]/T .
\eqno (9)
$$
The total energy per particle $E(T)$ is obtained from the energy density.
The excitation energy per particle is then calculated as
$E^{\star} = E(T)- E(T=0)$.

The continuum states of a nucleus at a nonzero temperature are occupied 
with a finite probability. The particle density therefore does not vanish
at large distances. The observables then depend on the size of the box
in which the calculations are done. In model calculations in heavy ion        
collisions, it is usually assumed that thermalisation occurs in a
freeze-out volume significantly larger than the normal nuclear volume.
Guided by this common practice, we fix a volume, find out the density
profiles and obtain the caloric curve, the excitation energy as a function
of temperature. The specific heat at constant volume is then calculated as
$ C_v = (dE^{\star}/dT)_v$.

Starting from the guess density, one can
arrive at a self-consistent density iteratively and calculate the physical
observables. The details are given in \cite{DE}.
\medskip

{\bf 2.3. Inclusion of collective radial flow}
\vskip 0.5cm
A hot nuclear system created in the laboratory from energetic nuclear
collisions may be initially compressed. In the decompression phase,
a collective radial flow may be generated in addition to thermal
excitation. An expanding system, in a strict thermodynamic sense is not
in equilibrium. However, if the time scale involved in the expansion is
much larger compared to the equilibration times in the expanding complex,
the thermodynamic equilibrium concept may still be meaningful. 
In a recent paper, Pal et al.
\cite{PAL} suggested that the effect of collective radial flow could be
simulated through the inclusion of an external negative pressure in the
total thermodynamic potential at freeze-out volume. 
A positive uniform external
pressure gives rise to compression; similarly a negative pressure gives
rise to an inflationary scenario resulting in the outward radial flow of
matter. The expanding system can hence be assumed to be under the action
of a negative external pressure $P_0$. Its magnitude is equal to flow    
pressure $P_f (|P_0| = P_f)$, the internal pressure exerted by the radially
outgoing nucleons at the freezeout surface. The total thermodynamic 
potential of the system at freezeout then is given by
$$
G=E-TS-\sum_{\tau}\mu_{\tau}A_{\tau} + P_0\Omega,
\eqno (10)
$$
where $P_0$ is the constant external pressure assumed negative and
$\Omega$ the effective volume. It is given as
$$
\Omega=\frac{4}{3}\pi R^3\,,\,\,\,\,\,R=(\frac{5}{3}<r^2>)^{1/2}.
\eqno (11)
$$
Here $<r^2>^{1/2}$ is the rms radius of the matter density distribution
and R is the radius of the corresponding uniform density distribution.
The occupancy obtained from the minimisation of the thermodynamic
potential then contains a pressure term and reads as
$$
n_{\tau}(\vec r,\vec p)=[1+exp\{(\frac{p^2}{2m_{\tau}^{\star}(r)}+V_{\tau}^0(r) +
V_{\tau}^2(r)+\delta_{\tau p}V_c(r)-\mu_{\tau} + 
P_0\frac{10\pi}{3A}Rr^2)/T\}]^{-1}.
\eqno (12)
$$
With the inclusion of flow, the whole set of calculations can then
be repeated with an effective chemical potential
$$
\mu_{\tau}^{eff}=\mu_{\tau}-P_0\frac{10\pi}{3A}Rr^2.
\eqno (13)
$$
It has been shown \cite{PAL} that the flow pressure $P_f(=-P_0)$ is
related to the flow energy as
$$
P_f=D(v_f,T)\rho (r)e_f(r),
\eqno (14)
$$
where the quantity $D(v_f,T)$ depends weakly on the temperature T
and the radially directed flow velocity $v_f$ and is $\simeq$ 4.5
for nucleons and $e_f(r)$ the flow energy per nucleon at any point
within the volume. The total flow energy may be expressed as 
$$
E_f= \int_{v} \rho(r) e_f(r) d{\vec r}
=P_f V/D(v_f,T).
\eqno (15)
$$ 
Here $\int_{v}=V$ is the size of the box or the freeze-out volume in
which the calculations are done.
\vskip 0.5cm
3. RESULTS AND DISCUSSIONS
\vskip 0.5cm
To do calculations in the mass range of experimental interest,
we choose two systems, namely, $^{150}$Sm and $^{85}$Kr. We report
calculations done in a freeze-out volume $V=8V_0$ where $V_0$ is
the normal volume of the nucleus at zero temperature. The freeze-out
volume chosen here is consistent with most multifragmentation
calculations \cite{GRO} and we have found that our results are
nearly insensitive to volumes much beyond these chosen confinement
volumes. Initially, we report calculations without any flow
effects. 
In Figure 1, the proton density for $^{150}$Sm
is displayed at four temperatures, T= 5 MeV (long-dashed curve), T= 9 MeV
(dotted curve), T= 9.5 MeV (dash curve) and T= 10 MeV (full
curve). With increasing temperature, the central density is 
progressively depleted with thickening of the long tail spread at the
boundary. Beyond T = 9.5 MeV, the density change is rather abrupt and
at T close to 10 MeV, the whole system looks like a uniform distribution
of matter inside the volume. This is shown by a representative density 
distribution at T = 10 MeV. The slight bump at the edge of the density
is due to the Coulomb force.
\begin{figure}
\centerline{\psfig{figure=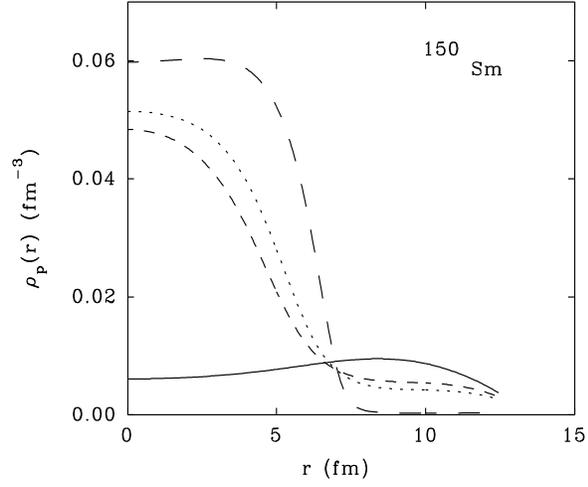,height=2.5in,width=3in}}
\caption{The proton density profile for the system $^{150}Sm$ calculated
at four temperatures as mentioned in the text.}
\end{figure}
The caloric curve for the system $^{150}$Sm is shown in Figure 2 
(dotted line). Initially the temperature rises faster with excitation
energy, then the rise is slower. 
\begin{figure}
\centerline{\psfig{figure=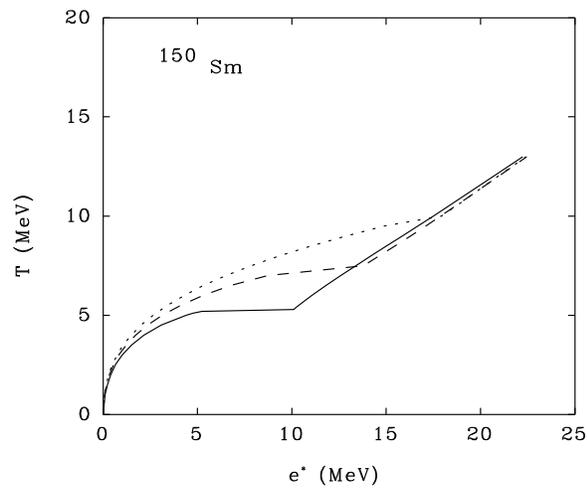,height=2.5in,width=3in}}
\caption{The caloric curve for the system $^{150}Sm$ at three pressures.}
\end{figure}
A kink is observed at $T\approx$10
MeV, beyond which the excitation energy rises linearly with temperature.
The slope $C_v$ is then 3/2 which is a reflection of the fact that
the nucleons then behave almost like a classical gas of particles. 
In Figure 3, the specific heat at constant volume $C_v$ is displayed
(dotted line). It shows a sharp peak at a temperature $T \simeq 10$
MeV. We believe this to be possibly a signal of the liquid-gas phase
transition. The abrupt transition in the density
at this temperature and the sharp fall of $C_v$ to the classical value
of 3/2 lend credence to our surmise.
\begin{figure}
\centerline{\psfig{figure=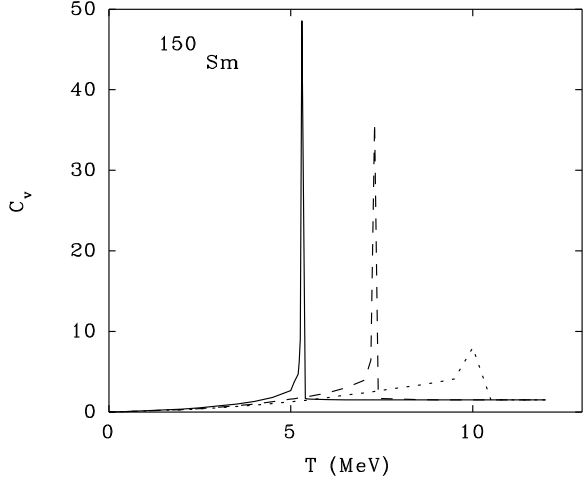,height=2.5in,width=3in}}
\caption{The specific heat per particle plotted as a function
of temperature for $^{150}Sm$. The three curves refer to
$P_0=0.0$ (dotted), $P_0=-0.05$ (dashed) and $P_0=-0.1$ (full curve)
respectively.}
\end{figure}
Collisions between heavy nuclei at high energies may generate a 
modest amount of compression even for far-central impacts. 
\begin{figure}
\centerline{\psfig{figure=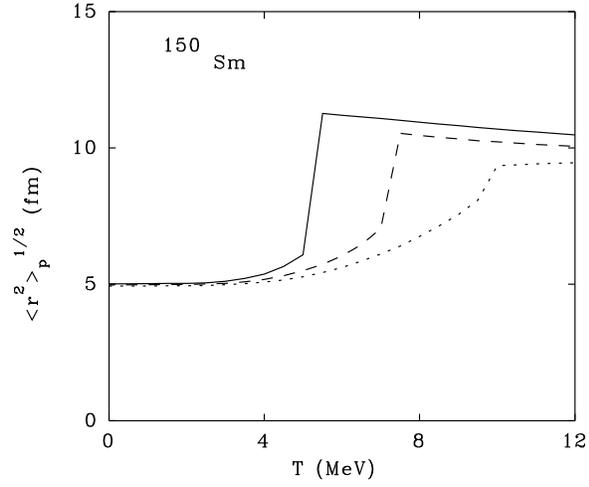,height=2.5in,width=3in}}
\caption{The proton rms radius as a function of temperature 
for $^{150}Sm$ at three pressures.}
\end{figure}
We have therefore repeated the calculations taking into account
the effect of flow energy with different values of $P_0$. If
$P_0$ is given, an estimate of the flow energy per nucleon
can be easily made with the help of Equations (14) and (15).
For example, if $P_0= -0.1$ MeV $fm^{-3}$, the average flow 
energy per nucleon is $\simeq 1.3$ MeV. The caloric curves for
$P_0=-0.05$ (dashed line) and $-0.1$ MeV $fm^{-3}$ (full line) are displayed
also in Figure 2. We find that with increase in flow energy,
the rise in temperature is slower and when the pressure $P_0=-0.1$
MeV $fm^{-3}$, the caloric curve shows a plateau at $T \simeq 5$
MeV in the excitation energy range of 5-10 MeV. In Figure 3, the
corresponding specific heats are displayed. The broad-based peak
for no flow goes over to an extremely sharp peak from $T \simeq 10$
MeV to $T \simeq 5$ MeV
with increasing flow. Looking at Figures
2 and 3, it is evident that the system signals a liquid-gas phase
transition at the peak temperatures and that the system moves
from a continuous phase transition to a sharp first order phase
transition. 
\begin{figure}
\centerline{\psfig{figure=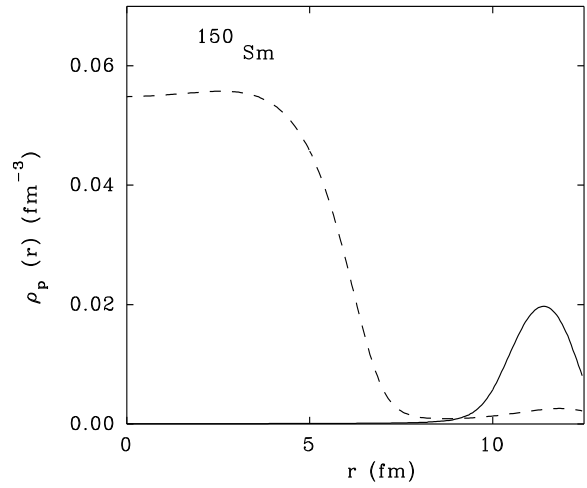,height=2.5in,width=3in}}
\caption{The proton density profile for 
$^{150}Sm$ with $P_0=-0.1$ MeV~fm$^{-3}$ for $T=5.2$ MeV (dashed line)
and $T=5.3$ (full line) MeV.}
\end{figure}
With increase in temperature, it is natural to expect that the rms
radius of the nucleus in question would increase. That is displayed
in Figure 4. 
Interestingly, we find that for cases accompanied with
flow energy, there is an extremely sharp increase in the rms radius
at the phase transition temperature. From Figure 5, we notice that
near the phase transition temperature, within an interval of 
$\Delta T \approx 0.1$ MeV, the density disribution now undergoes
an exotic shape transition to a bubble shape at T= 5.3 MeV.
This may be attributed to the loss of surface
tension at the transition temperature; when there is flow, the flow 
pressure pushes the  matter outwards which takes a bubble shape 
because of the constraint of a spherical freeze-out volume. The
calculations have been repeated for $^{85}$Kr with freeze-out 
volume 8 times its normal volume and all the features reported
for $^{150}$Sm are reproduced except for the fact that the transition
temperature is shifted up by $\approx$ 0.5 MeV and the peak for the
specific heat is broader.
\vskip 0.5cm
4. CONCLUSIONS
\vskip 0.5cm
We have calculated the caloric curve and the specific heat of finite
nuclei in a self-consistent finite temperature Thomas-Fermi theory
with a realistic effective nucleon-nucleon interaction. The effects of
collective radial flow, if any, are built in the theoretical framework.
>From the caloric curve and the specific heat, signals of liquid-gas
phase transition are obtained for cases both with and without collective
flow. In the case without flow, there appears to be a continuous phase
transition at a temperature $T \simeq 9.5$ MeV whereas with inclusion 
of a little flow energy, the phase transition seems to be of first order
at a temperature $T \simeq 5.0$ MeV. The nucleus also then changes its
shape from a diffuse sphere to a bubble. A qualitative reference to
the two sets of experimental data referred to in the introduction 
may not be out of place. In the Au induced reaction on C at 1 AGeV,
for near peripheral collisions, there is possibly little or no
compression because of the small target size. The caloric curve then
alludes to a continuous phase transition as theoretically obtained
and experimentally reported. For the Au on Au reaction at 600 AMeV, 
a modest compression and hence radial flow may not be ignored and then
a sharper phase transition may result as indicated in the experiments.

This work is partially supported by the U.S. National Science Foundation
under grant no. PHY-9413872.

\end{document}